\documentclass[aps,pra,twocolumn,groupaddress,10pt]{revtex4-1} %maybe a better one?
\usepackage{amsmath}
\usepackage{latexsym}
\usepackage{graphicx}
\usepackage{amsfonts}
\usepackage{braket}
\usepackage{hyperref}
\usepackage{natbib}
\usepackage{amssymb}
\usepackage{mathtools} %for rcases
\usepackage{nicefrac}
\usepackage{fancyhdr}
\usepackage[utf8]{inputenc}
\usepackage{dsfont}
\usepackage{colortbl}
\usepackage{xcolor}
\usepackage{subfigure}
\usepackage{psfrag}
\usepackage{tikz}
\usepackage{paralist}
\usepackage{nicefrac}
\usepackage{ulem}
\usetikzlibrary{arrows,shapes,decorations.pathmorphing}

\hypersetup{
    bookmarks=true,         % show bookmarks bar?
    bookmarksopen=true,			% expand outlines in pdf at start?
    bookmarksopenlevel=1,		% level up to which is expanded
    unicode=false,          % non-Latin characters in Acrobat’s bookmarks
    pdftoolbar=true,        % show Acrobat’s toolbar?
    pdfmenubar=true,        % show Acrobat’s menu?
    pdffitwindow=false,     % window fit to page when opened
    pdfstartview={FitH},    % fits the width of the page to the window
    pdftitle={Non-adiabatic corrections to fast dispersive multiqubit gates involving Z-control},    % title
    pdfauthor={Lukas Theis},     % author
    colorlinks=true,       % false: boxed links; true: colored links
    linkcolor=blue,          % color of internal links (change box color with linkbordercolor)
    citecolor=blue,        % color of links to bibliography
    urlcolor=blue           % color of external links
}

	\newcommand{\comut}[2]{[ #1 , #2 ]} % THE COMMUTATOR
	\newcommand{\Comut}[2]{\left[ #1 , #2 \right]} % THE COMMUTATOR
	\newcommand{\op}[1]{#1}

	\newcommand{\erf}[1]{\mathrm{erf}\left(#1\right)}
	
	\newcommand{\real}[1]{\operatorname{Re}\left\{#1\right\}} %real part
	\newcommand{\imag}[1]{\operatorname{Im}\left\{#1\right\}} %imaginary part
	\renewcommand{\a}{\op{a}}
	\newcommand{\ad}{\op{a}^{\dagger}}
	\newcommand{\sm}[2]{\op{\sigma}_{#2}^{-(#1)}}
	\renewcommand{\sp}[2]{\op{\sigma}_{#2}^{+(#1)}}
	
	\renewcommand{\cos}[1]{\mathrm{cos}\!\left(#1\right)}
	\renewcommand{\sin}[1]{\mathrm{sin}\!\left(#1\right)}
	\renewcommand{\tan}[1]{\mathrm{tan}\left(#1\right)}
	\renewcommand{\exp}[1]{\mathrm{exp}\left(#1\right)}
	\renewcommand{\erf}[1]{\mathrm{Erf}\left(#1\right)}
	\newcommand{\bigchi}{\protect\scalebox{0.8}{\raisebox{1pt}{$\chi$}}}% big chi for fractions in matrix

	\newcommand{\del}{\mathrm{\Delta}}
	\newcommand{\ketbra}[2]{\ket{#1}\!\bra{#2}}

	\newcommand{\LR}[1]{\left(#1\right)}

	\newcommand{\Abs}[1]{\left\lvert #1 \right\rvert}

	\newcommand{\TD}[2]{\frac{\text{d}}{\text{d} #2}\left(#1\right)}
	\newcommand{\Trace}[1]{\text{Tr}\LR{#1}}		% Trace
	\newcommand{\reffig}[1]{Fig.\,\ref{fig:#1}}
	\renewcommand{\refeq}[1]{(\ref{eq:#1})}

\makeindex

\begin{document}

\title{Non-adiabatic corrections to fast dispersive multiqubit gates involving $Z$-control}

\author{L. S. Theis}\email{luk@lusi.uni-sb.de}\author{F. K. Wilhelm}
\affiliation{Theoretical Physics, Saarland University, 66123 Saarbr{\"u}cken, Germany}

\makeatletter
\def\Dated@name{Date: }
\makeatother
\date{\today}

\begin{abstract}
We review a time-dependent version of the Schrieffer-Wolff transformation that accounts for real-time control of system parameters, soon to be rendered possible on a broad basis due to technical progress. The dispersive regime of $N$ multilevel systems coupled to a  cavity via a Jaynes-Cummings interaction is extended to the most general case. As a concrete example we rigorously apply the technique to dispersive two-qubit gates in a superconducting architecture, showing that fidelities based on previous models are off by up to $10^{-2}$, which is certainly relevant for high-fidelity gates compatible with fault-tolerant quantum information devices. A closed analytic form for the error depending on the target evolution closes our work.
\end{abstract}

\maketitle

% -----------------------------------------------------------------------------------
\section{Introduction}
In many branches of physics and other natural sciences, analyzing dynamics of certain systems of interest can often be vastly simplified if it is possible to separate time scales, such as it is possible for a spinning top: it spins at a high frequency whereas its precession frequency is usually much lower. It is then often advantageous to apply frame transformations that separate the subspace of interest from the rest, such as separating a low-energy(frequency) subspace from a high-energy(frequency) subspace. A prominent and well-celebrated technique in quantum physics is the Schrieffer-Wolff transformation \cite{Bravyi_Annals_326_2793} which is named after the authors of a famous condensed matter paper \cite{Schrieffer_PhysRev_149_491} that relates the Anderson Hamiltonian to the Kondo Hamiltonian. In fact, the transformation has already been used multiple times many years before -- for instance in order to study the dynamics of rotating molecules \cite{Vleck_PhysRev_33_467}, which is why the method is also known as \emph{van Vleck perturbation theory}. To our knowledge, the first application in quantum physics was about 15 years before Ref. \cite{Schrieffer_PhysRev_149_491} in Foldy's and Wouthuysen's work about the Dirac theory of spin 1/2 particles \cite{Foldy_PhysRev_78_29}. However, for convenience, we will refer to the technique as Schrieffer-Wolff transformation (SWT). A related method, so called \emph{adiabatic perturbation theory} \cite{Teufel_perturbation,Rigolin_PRA_78_052508}, perturbatively extends the adiabatic approximation in order to solve the effective dynamics of Hamiltonians that feature such a separation of scales.

Currently, applications of the SWT are countless. Apart from the examples mentioned before, it is widely used in quantum many-body systems. The SWT can for instance be used to study electron gases \cite{Erlingsson_PRB_82_155456} and the ground state of the Hubbard model \cite{Oles_PRB_41_2562}, but it has also become an important tool in quantum information theory. For instance, it aids the understanding of the dispersive interaction in circuit quantum electrodynamics \cite{Blais_PRA_75_032329, Govia_PRA_93_012316, Khezri_PRA_94_012347} within the framework of superconducting qubits \cite{Clark_Nature_453_1031} coupled through a resonator \cite{DiCarlo_Nature_460_240}. 

An important property of the SWT is that the eigenvalues of the derived effective Hamiltonian reproduce those of the full Hamiltonian (in the relevant subspace) to the required approximation. It may happen that the derived effective Hamiltonian has fewer  degrees of freedom than the full Hamiltonian while featuring a more complex structure, which has eventually inspired the idea of perturbation gadgets \cite{Kempe_2006,Oliveira_2008}, where the SWT is used to analyze and construct high-energy simulator Hamiltonians with a low degree of complexity that are used to approximate complex low-energy dynamics of some target Hamiltonian \cite{Bravyi_PRL_101_070503}.

With ongoing technical developments, real-time control of quantum systems has become an important tool in quantum information to assess new degrees of controllability. However, applications of real-time control are not only  limited to quantum information processing \cite{Cottet_PRB_86_075107}. Possible examples in the field of quantum information are quantum quenches in many-body systems, where a Hamiltonian is suddenly changed non-adiabatically \cite{Abeling_PRB_93_104302}, or fast tuning of qubit frequencies \cite{Martinis_PRA_90_022307,Chasseur_PRA_31_043421}. Frequency-tuning of superconducting qubits is typically done by changing the magnetic flux penetrating the Josephson junctions \cite{Strand_PRB_87_220505}. This method is quite sensitive to flux noise, which is why fast real-time flux control has so far been a difficult task. Yet, recent development of a new qubit design \cite{Lange_PRL_115_127002}, called the \emph{Gatemon}, allows for fast frequency-tuning by manipulating voltage \cite{Casparis_PRL_116_150505} instead of magnetic flux, so that fast frequency sweeps are easily possible. 

In this work, we briefly review the idea of the SWT and present a general extension of the method incorporating time-dependent effects, which is inevitable given the imminent implementation of real-time controls. In Ref.  \cite{Goldin_PRB_61_16750} Goldin and Avishai have used a time-dependent analogue of the SWT to study time-dependent impurities in Anderson and Kondo models. We adapt the idea of constructing a time-dependent Schrieffer-Wolff transformation (TDSWT) and present the full hierarchy of the approximation. Performing a second order perturbation theory ultimately reveals that the TDSWT adiabatically eliminates terms in the Hamiltonian that originate in real-time control. A similar idea of frame transformations has been used to adiabatically eliminate leakage errors in anharmonic ladder systems \cite{Motzoi_PRL_103_110501}, such as superconducting qubits, but has -- in a generalized version -- for instance also proven promising to reduce errors in Rydberg gates \cite{Theis_PRA_94_032306}. By way of example we reconsider the dispersive transformation of a Jaynes-Cummings type Hamiltonian for arbirarily many multilevel systems, taking into account that the energy levels as well as the couplings in general depend on external controls, such as e.g. magnetic flux for Transmon qubits \cite{Koch_PRA_76_042319}. 

% In this work, we briefly review the idea of the SWT and present a general extension of the method incorporating time-dependent effects \cite{Goldin_PRB_61_16750}, which is inevitable given the imminent implementation of real-time controls. We show that the time-dependent Schrieffer-Wolff transformation (TDSWT) adiabatically eliminates terms in the Hamiltonian that originate in real-time control. A similar idea of frame transformations has been used to adiabatically eliminate leakage errors in anharmonic ladder systems \cite{Motzoi_PRL_103_110501}, such as superconducting qubits, but has -- in a generalized version -- for instance also proven useful to reduce errors in Rydberg gates \cite{Theis_PRA_94_032306}. For our purpose, we examplarily reconsider the dispersive transformation of a Jaynes-Cummings type Hamiltonian for arbirarily many multilevel systems, taking into account that the energy levels as well as the couplings in general depend on external controls, such as e.g. magnetic flux for Transmon qubits \cite{Koch_PRA_76_042319}. 

We focus on a system that is relevant for the implementation of entangling gates with superconducting qubits. However, similar arguments hold for instance for a quantum dot architecture, where the couplings depend on the external laser controls \cite{Imam_PRL_83_4204}. To substantiate the importance of our work, we show that the difference in fidelities, based on previous models and our extended one, can be on the order of $10^{-2}$ which is of indisputable importance for high-fidelity gates, given that the error threshold for fault-tolerant quantum error correction is believed to lie between $10^{-4}$ and $10^{-2}$ for many relevant systems \cite{Devitt_2013}. A second-order Magnus expansion \cite{Warren_JChemPhys_81_5437,Blanes_PhysRep_470_151} provides a closed analytic form to accurately estimate the errors observed in numerically exact simulations.  

\section{The Schrieffer-Wolff transformation}
\subsection{Review of the original idea}
The essence of the original SWT \cite{Schrieffer_PhysRev_149_491} is to generate an effective Hamiltonian $\op{\tilde{H}}$ from the Hamiltonian
\begin{align} \label{eq:hamiltonian}
  \op{H}=\op{H_0}+\underbrace{\epsilon\LR{\op{H}_1+\op{H}_2}}_{\op{H}_I}
\end{align}
using a perturbative expansion, so that $\op{\tilde{H}}$ is block-diagonal up to a desired order in the perturbing term $\op{H}_I$. It is advantageous to separate the perturbation into a block-diagonal term $\op{H}_1$ and a block-offdiagonal one $\op{H}_2$. The effective Hamiltonian is obtained via a unitary transformation $e^{\op{S}}$ so that $\op{\tilde{H}}=e^{-\op{S}}\op{H}e^{\op{S}}$. Here, $\op{S}$ is an anti-hermitian operator (this also preserves the Lie structure of the problem \cite{Primas_RevModPhys_35_710}) and can be written as
\begin{align}\label{eq:swt1}
 \op{\tilde{H}} = e^{-\op{S}}\op{H}e^{\op{S}} = \sum\limits_{j=0}^{\infty}\frac{1}{j!}\comut{\op{H}}{\op{S}}_{j},
\end{align}
whereby $\comut{A}{B}_{n}=\comut{\comut{A}{B}_{n-1}}{B}$ and $\comut{A}{B}_{0}=A$. A typical, but not mandatory way to determine the sought transformation is to expand $\op{S}$ in different orders of $\epsilon$, i.e. $\op{S}=\sum_j \op{S}_j$, which ultimately allows one to remove the off-diagonal perturbation $\op{H}_2$ up to a desired order in $\epsilon$. One obtains successive equations for the $\op{S}_j$ from an order-by-order expansion in the perturbation $\epsilon$, e.g. $\comut{\op{H}_0}{\op{S}_1}=-\op{H}_2$ removes the off-diagonal perturbation up to lowest order. More details are provided in the following section which extends the SWT to a generic time-dependent case.\\[.5em]
\subsection{Extension to time-dependent perturbations}
The formalism of the previous section needs to be extended \cite{Goldin_PRB_61_16750}, as soon as the perturbing term is time-dependent. Then, the operator $\op{S}$ in general is time-dependent so that the transformation in Eq.\refeq{swt1} needs to be extended to
\begin{align}\label{eq:swt2}
 \op{\tilde{H}}=e^{-\op{S}}\op{H}e^{\op{S}}+i\partial_{t}\LR{e^{-\op{S}}}e^{\op{S}}.
\end{align}
Analogously to before, Eq.\refeq{swt2} is expanded in terms of commutators so that the effective Hamiltonian can be written as
\begin{align}
 \op{\tilde{H}} & = \sum\limits_{j=0}^{\infty}\frac{1}{j!}\comut{\op{H}}{\op{S}}_{j}-i\sum\limits_{j=0}^{\infty}\frac{1}{(j+1)!}\comut{\dot{\op{S}}}{\op{S}}_{j}.
\end{align}
Due to the decomposition of Hamiltonian \refeq{hamiltonian} we can separate the effective Hamiltonian $\op{\tilde{H}}$ into off-diagonal ($\op{\tilde{H}}_{\rm od}$) and diagonal ($\op{\tilde{H}}_{\rm d}$) terms, which are given by
\begin{widetext}
 \begin{subequations}\label{eq:effective_hamiltonians}\begin{align}
  \op{\tilde{H}}_{\rm od} & = \sum\limits_{j=0}^{\infty}\frac{1}{(2j+1)!}\comut{\op{H}_0+\op{H}_1}{\op{S}}_{2j+1} + \sum\limits_{j=0}^{\infty}\frac{1}{(2j)!}\comut{\op{H}_2}{\op{S}}_{2j} - i\sum\limits_{j=0}^{\infty}\frac{1}{(2j+1)!}\comut{\dot{\op{S}}}{\op{S}}_{2j}\\
  \op{\tilde{H}}_{\rm d} & = \sum\limits_{j=0}^{\infty}\frac{1}{(2j)!}\comut{\op{H}_0+\op{H}_1}{\op{S}}_{2j} + \sum\limits_{j=0}^{\infty}\frac{1}{(2j+1)!}\comut{\op{H}_2}{\op{S}}_{2j+1} - i\sum\limits_{j=0}^{\infty}\frac{1}{(2j+2)!}\comut{\dot{\op{S}}}{\op{S}}_{2j+1}.\label{eq:effective_diagonal}
 \end{align}\end{subequations}
\end{widetext}
An expansion of $\op{S}=\sum_j\op{S}_j$ as a power series in the perturbation yields equations that solve $\op{\tilde{H}}_{\rm od}=0$ up to the desired order in $\epsilon$, and thereby block-diagonalize the Hamiltonian. As stated before, this choice for $\op{S}$ is not mandatory, but the typical choice for a perturbative expansion. Consequently different ans{\"a}tze for $\op{S}$ lead to different block-diagonalizations. In order to compare orders of $\epsilon$, we make the \emph{a priori} assumption that $\dot{\op{S}}_j$ is of order $j+1$ of the perturbation. Hence, the first few equations that determine the transformation read

 \begin{subequations}\label{eq:successive_commutator}\begin{align}
  \comut{\op{H}_0}{\op{S}_1}  =& -\op{H}_2 \label{eq:com1}\\
  \comut{\op{H}_0}{\op{S}_2}  =& -\comut{\op{H}_1}{\op{S}_1} + i\dot{\op{S}}_1 \label{eq:com2}\\
  \comut{\op{H}_0}{\op{S}_3}  =& -\comut{\op{H}_1}{\op{S}_2} -\frac{1}{3}\comut{\op{H}_2}{\op{S}_1}_{2} + i\dot{\op{S}}_2\label{eq:com3}\\
  \begin{split}\comut{\op{H}_0}{\op{S}_4}  =& -\comut{\op{H}_1}{\op{S}_3} - \frac{1}{3}\comut{\comut{\op{H}_2}{\op{S}_1}}{\op{S}_2}\\& - \frac{1}{3}\comut{\comut{\op{H}_2}{\op{S}_2}}{\op{S}_1} + i\dot{\op{S}}_3.\end{split}
 \end{align}\end{subequations}

Successively solving Eqs.\refeq{successive_commutator} will then cancel all perturbing terms up to the desired order so that the effective Hamiltonian $\op{\tilde{H}}$ is purely block-diagonal. We need to check the consistency of the solutions to Eqs.\refeq{successive_commutator} under the \emph{a priori} assumption on the derivative of $\dot{\op{S}}_j$: from Eq.\refeq{com1} we see that $\op{S}_1$ inherits perturbation of order one from $\op{H}_2$. Similarly, it follows from Eq.\refeq{com2} that $\dot{\op{S}}_1$ and $\op{S}_2$ are of order two in the perturbation and so on. This verifies the consistency of the expansion. Finally, the block-diagonal terms in Eq.\refeq{effective_diagonal} need to be calculated. Using Eqs.\refeq{successive_commutator} the block-diagonal contributions can be simplified so that the first few remaining terms that constitute the effective Hamiltonian $\op{\tilde{H}}=\sum_j\op{\tilde{H}}_j$ read
\begin{subequations}\label{eq:solEffHam}\begin{align}
 \op{\tilde{H}}_0  =& \op{H}_0\\
 \op{\tilde{H}}_1  =& \op{H}_1\\
 \op{\tilde{H}}_2  =& \frac{1}{2!}\comut{\op{H}_2}{\op{S}_1}\\
 \op{\tilde{H}}_3  =& \frac{1}{2!}\comut{\op{H}_2}{\op{S}_2}\\
 \op{\tilde{H}}_4  =& \frac{1}{2!}\comut{\op{H}_2}{\op{S}_3}-\frac{1}{4!}\comut{\op{H}_2}{\op{S}_1}_3
\end{align}\end{subequations}

\subsection{The dispersive transformation}
A particular example of the SWT in the context of quantum information is the analysis of cavity-mediated residual interactions between multilevel systems. Under the assumption of weak coupling, the SWT can be used to derive an effective Hamiltonian which is free of interactions between multilevel systems and the cavity. Many fundamental concepts, such as readout \cite{Blais_PRA_69_062320} and gate synthesis \cite{Blais_PRA_75_032329}, are based on this so called \emph{dispersive frame}. With the notation for the Hamiltonians adopted from the previous sections and the convention that $\hbar=1$, we write the Jaynes-Cummings Hamiltonian \cite{Jaynes_IEEE_51} for $N$ multilevel systems with energy levels $\omega_{j}^{(m)}$ coupled to a cavity with coupling strengths $g_{j,j+1}^{(m)}$ as 
\begin{subequations}\label{eq:jcham}\begin{align}
  \op{H}_0 & =  \omega_r\ad\a + \sum\limits_{m=0}^{N}\sum\limits_{j=0}^{\infty}\omega_{j}^{(m)}\ketbra{j}{j}\\
  \op{H}_1 & = 0\\
  \op{H}_2 & = \sum\limits_{m=1}^{N}\sum\limits_{j=0}^{\infty}g_{j,j+1}^{(m)}\LR{\sp{m}{j}\a+\sm{m}{j}\ad}
\end{align}\end{subequations}
where superscript $(j)$ labels the $j$th element in the total Hilbert space and $\op{\Pi}_j^{(m)}\equiv\ketbra{j}{j}^{(m)}$. For readability, we abstain from highlighting time-dependent parameters, but want to remind the reader that the $\omega_j^{(m)}=\omega_j^{(m)}(t)$ and the couplings $g_{j,j+1}^{(m)}=g_{j,j+1}^{(m)}(t)$ are in general time-dependent quantities. The raising and lowering operators of each multilevel system, $\sp{m}{j}$ and $\sm{m}{j}$, are defined as 
\begin{subequations}\begin{align}
 \sp{m}{j} = & \ketbra{j+1}{j}^{(m)},\\
 \sm{m}{j} = & \ketbra{j}{j+1}^{(m)}.
\end{align}\end{subequations}
We aim at removing all interactions between the cavity and the multilevel systems up to second order, so that the dynamics can be solely reduced to the multilevel systems. Therefore, we need to find the operators $\op{S}_1$ and $\op{S}_2$ that satisfy Eqs.\refeq{com1} and \refeq{com2} for the Hamiltonians given by Eqs.\refeq{jcham}. Using
\begin{align}
 \Comut{\sum\limits_{j=0}^{\infty}\LR{\sm{m}{j}\ad\pm\sp{m}{j}\a}}{\op{H}_0} \propto \sum\limits_{j=0}^{\infty}\LR{\sm{m}{j}\ad\mp\sp{m}{j}\a}
\end{align}
solutions to $\op{S}_{1,2}$ are found. With the shorthand notation $\del_{j}^{(m)}\equiv\omega_{j,j+1}^{(m)}-\omega_{r}$ and $\omega_{j,j+1}^{(m)}\equiv\omega_{j+1}^{(m)}-\omega_{j}^{(m)}$ we write the corresponding solutions as
\begin{subequations}\begin{align}
 \op{S}_1 = & \hphantom{-i}\sum\limits_{m=1}^{N}\sum\limits_{j=0}^{\infty} \frac{g_{j,j+1}^{(m)}}{\del_{j}^{(m)}}\LR{\sm{m}{j}\ad-\sp{m}{j}\a},\\
 \op{S}_2 = & -i\sum\limits_{m=1}^{N}\sum\limits_{j=0}^{\infty} \frac{1}{\del_j^{(m)}}\TD{\frac{g_{j,j+1}^{(m)}}{\del_j^{(m)}}}{t}\LR{\sm{m}{j}\ad+\sp{m}{j}\a}.
\end{align}\end{subequations}
The so called \emph{dispersive Hamiltonian} up to second order then reads 
\begin{widetext}\begin{align}\begin{split}\label{eq:dispHamFull}
\op{\tilde{H}} = &\left\{\omega_{r} + \sum\limits_{m=1}^{N}\sum\limits_{j=1}^{\infty}\LR{\chi_{j-1,j}^{(m)}-\chi_{j,j+1}^{(m)}}\op{\Pi}_j^{(m)} - \sum\limits_{m=1}^{N}\chi_{0,1}^{(m)}\op{\Pi}_0^{(m)}\right\}\ad\a%
+ \sum\limits_{m=1}^{N}\omega_0^{(m)}\op{\Pi}_0^{(m)} + \sum\limits_{m=1}^{N}\sum\limits_{j=1}^{\infty}\LR{\omega_{j}^{(m)}+\chi_{j-1,j}^{(m)}}\op{\Pi}_j^{(m)}\\
& + \sum\limits_{m\neq n}\sum\limits_{j,k=0}^{\infty} g_{j,j+1}^{(m)}\lambda_k^{(n)}\LR{\sm{m}{j}\sp{n}{k}+\sp{m}{j}\sm{n}{k}}%
 + i\sum\limits_{m\neq n}\sum\limits_{j,k=0}^{\infty} g_{j,j+1}^{(m)}\frac{\dot{\lambda}_k^{(n)}}{\del_k^{(n)}}\LR{\sm{m}{j}\sp{n}{k}-\sp{m}{j}\sm{n}{k}}.
\end{split}\end{align}\end{widetext}
Here we have denoted one of the expansion parameters as $\lambda_j^{(m)}\equiv g_{j,j+1}^{(m)}/\del_j^{(m)}$, which we will refer to as the \emph{dispersive parameter}, and introduced the dispersive shift
\begin{align}	
 \chi_{j,j+1}^{(m)}\equiv \frac{\LR{g_{j,j+1}^{(m)}}^2}{\del_j^{(m)}}.
\end{align}
The contribution from $\op{S}_2$ adiabatically eliminates a time-dependent qubit-cavity interaction that would be apparent if the usual SWT was applied and the effective Hamiltonian is then extended by the summand $i\partial_{t}\LR{e^{-\op{S}}}e^{\op{S}}$, describing inertial forces when the new frame is interpreted as an accelerated reference frame. In fact, Hamiltonian \refeq{dispHamFull} is almost identical to the commonly used dispersive Hamiltonian \cite{Boissonneault_PRA_79_013819}: The multilevel systems are energy-shifted by the dispersive shifts and are dispersively coupled  via $\sigma^+\sigma^-$ interactions to each other through the cavity, whereby the interaction strength scales as $1/\del$. Additionally, we observe the usual shift of the resonator frequency $\omega_r$ by a value that depends on the state of the multilevel systems, which ultimately can be used for readout purposes \cite{Govia_PRA_92_022335}. However, the dispersive coupling in the Hamiltonian \refeq{dispHamFull} obtained via TDSWT has an additional contribution (imaginary and different signs) that scales proportionally to $\dot{\lambda}_k^{(n)}$ -- essentially the speed at which the parameters are modulated. 

In order that our perturbative expansion which leads to Hamiltonian \refeq{dispHamFull} is valid we need to limit the magnitude of the expansion parameters. They need to meet the conditions 
\begin{subequations}\label{eq:dispCond}\begin{align}
 \lambda_j^{(m)} & \ll 1,\label{eq:cond1}\\
  \dot{\lambda}_j^{(m)}/\del_j^{(m)} & \ll 1 \label{eq:cond2}
\end{align}\end{subequations}
for all values of $j$ and $m$. Otherwise higher-order terms in Eqs.\refeq{solEffHam} need to be considered, which is straightforward and does not qualitatively change the results. Since Eq.\refeq{cond2} basically limits the velocity at which $\lambda$ may change, we refer to $\dot{\lambda}_j^{(m)}/\del_j^{(m)}$ as \emph{dispersive adiabaticity parameter}.\\[.5em]
\section{Example: Two Transmon qubits}
\subsection{Implementation of entangling gates}
As an example, we choose to work with two Transmon qubits \cite{Koch_PRA_76_042319} coupled to the same resonator $\omega_r$. From Eq.\refeq{dispHamFull} we see that the physical qubits dispersively couple to each other through the cavity. This interaction provides a common way to implement a controlled-phase gate: The avoided crossing between the $\ket{11}$ and $\ket{20}$ states can be used to control the phase of the $\ket{11}$ state \cite{DiCarlo_Nature_460_240}. Optimal control has sought fast pulses to produce high-fidelity gates based on this interaction using a geometric derivation \cite{Martinis_PRA_90_022307} as well as a deeper analysis of the underlying Landau-Zener physics \cite{Chasseur_PRA_31_043421}. For convenience, we only work in the relevant $\{\ket{11},\ket{20}\}$ subspace, where the reduced Hamiltonian $\op{\tilde{H}}_{\rm red}$ is given by
\begin{widetext}\begin{align}\label{eq:dispHamRed}
  \op{\tilde{H}}_{\rm red} & =
  \begin{pmatrix}
    \frac{\bigchi_t+\delta\omega-\alpha^{(1)}}{2} & i\LR{\frac{\dot{\lambda}_0^{(2)}g_{1,2}^{(1)}}{\del_0^{(2)}}-\frac{\dot{\lambda}_1^{(1)}g_{0,1}^{(2)}}{\del_1^{(1)}}}+g_{0,1}^{(2)}\lambda_1^{(1)}+g_{1,2}^{(1)}\lambda_0^{(2)}\\
    -i\LR{\frac{\dot{\lambda}_0^{(2)}g_{1,2}^{(1)}}{\del_0^{(2)}}-\frac{\dot{\lambda}_1^{(1)}g_{0,1}^{(2)}}{\del_1^{(1)}}}+g_{0,1}^{(2)}\lambda_1^{(1)}+g_{1,2}^{(1)}\lambda_0^{(2)} & -\frac{\bigchi_t+\delta\omega-\alpha^{(1)}}{2}
  \end{pmatrix}.
\end{align}\end{widetext}
Here we denote the anharmonicity of the first Transmon as $\alpha^{(1)}$ and use the definitions $\delta\omega=\omega_1^{(2)}-\omega_1^{(1)}$ and $\chi_t=\chi_{0,1}^{(1)}+\chi_{0,1}^{(2)}+\chi_{1,2}^{(1)}$. In previous implementations of two-qubit gates \cite{DiCarlo_Nature_460_240}, one qubit (Q1) is held at a constant frequency whereas the frequency of qubit two (Q2) changes in time: First being far detuned from Q1 and $\omega_r$, it is tuned down to a constant frequency close to Q1 to generate a strong dispersive interaction, interacts for a certain time $T$ and is tuned back from close-resonance again as soon as the interaction time $T$ was long enough to implement the desired gate. However, gate generation can be tremendously sped up by real-time control of frequencies via modulating the applied magnetic flux $\Phi$ \cite{Martinis_PRA_90_022307,Chasseur_PRA_31_043421}. The qubit frequencies $\omega_j$ as well as the couplings $g_{j,j+1}$ scale with the applied flux, which changes the Josephson energy $E_J$ of the Josephson junctions, as \cite{Koch_PRA_76_042319}
\begin{align}
	E_J(\Phi) & = E_{J\Sigma}\cos{\frac{\pi\Phi}{\Phi_0}}\sqrt{1+d^2\tan{\frac{\pi\Phi}{\Phi_0}}}\\
	g_{j,j+1}(\Phi) & \propto \sqrt{\frac{j+1}{2}}\LR{E_J(\Phi)}^{1/4}\label{eq:flux_couple}\\
	\omega_j(\Phi) & = j\sqrt{8E_cE_J(\Phi)}+\alpha_j,
\end{align}
whereby $\Phi_0$ is the magnetic flux quantum, $E_c$ the charging energy and $d$ is the junction asymmetry. Without loss of generality we focus on symmetric junctions, i.e. $d=0$. The anharmonicities $\alpha_j$ in case of a Duffing oscillator -- which is a good approximation for Transmon qubits -- are given by \cite{Gambetta_PRA_83_012308} the relation
\begin{align}
	\alpha_j & = \frac{j(j-1)}{2}\alpha_2.
\end{align}
\subsection{Time-dependent effects}
Since the qubit frequencies depend on time (flux control), so do the detunings $\del_j$. Moreover, the coupling strengths $g_{j,j+1}$ also depend on the applied flux as given by Eq.\refeq{flux_couple}. It is crucial to note that especially the flux-dependence of $g$ is usually not considered, and the effective Hamiltonians are derived without taking into account the effect of real-time parameter control in the SWT \cite{McKay_arXiv_1604.03076}. However, we will show that it is inevitable to rigorously incorporate the effect of $g$ and $\del$ changing with flux if one aims at high-fidelity gates, compatible with current error thresholds. Along these lines, it is also important to question the assumption of constant off-diagonal elements in the Hamiltonian \refeq{dispHamRed}, as for instance done in Refs. \cite{Martinis_PRA_90_022307,Chasseur_PRA_31_043421}. 

\begin{figure}
 \centering
 \includegraphics[width=.98\linewidth]{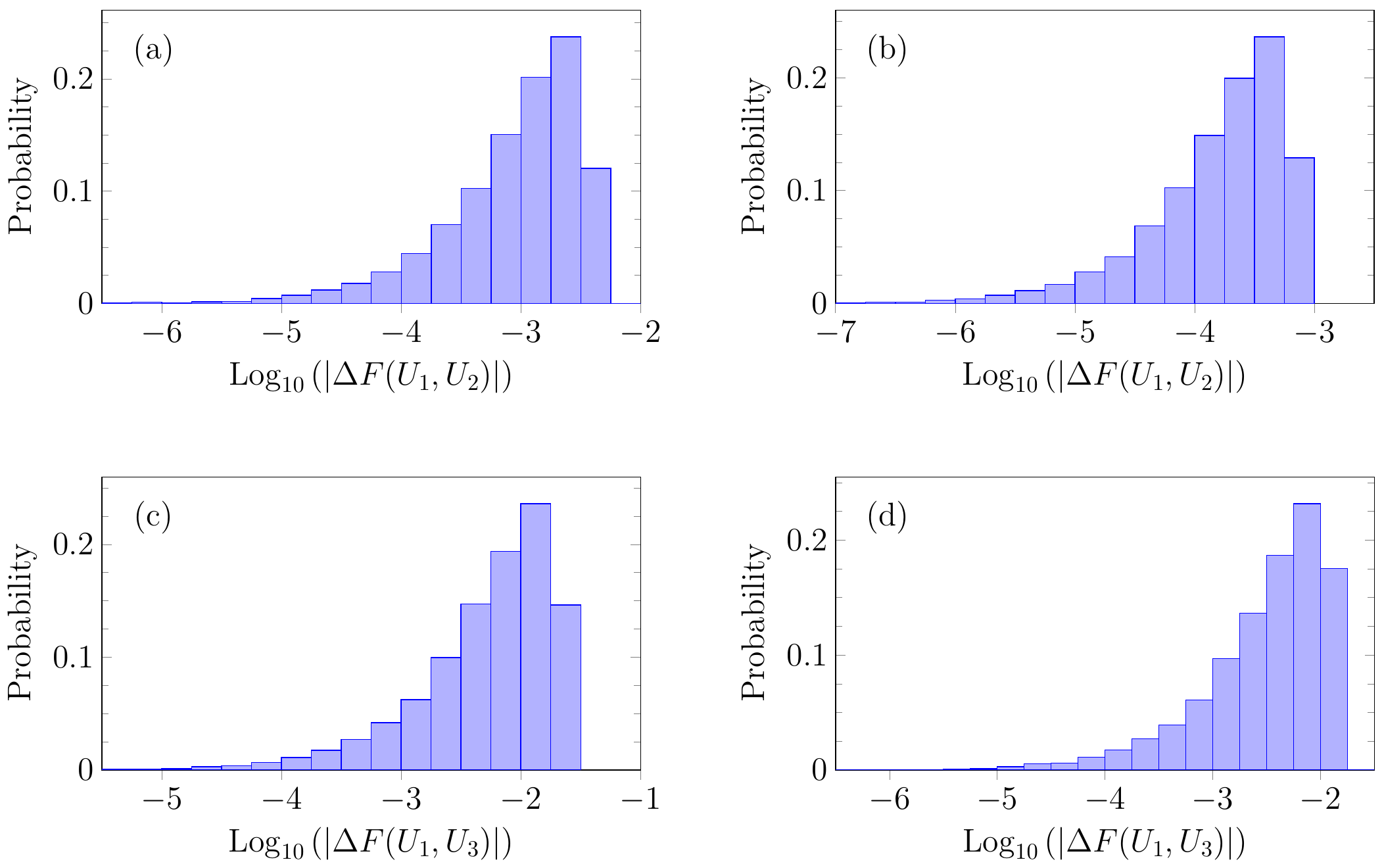}
 \caption{\label{fig:diff1}Normalized histograms for the differences in fidelity $\del F$ with respect to $N_s=10000$ random unitaries. Neglecting the time-dependent part of the dispersive transformation ($\propto \dot{\lambda}$) leads to errors on the order of $10^{-3}$ (top row), which is certainly relevant for high-fidelity gates. Models as were used before \cite{Martinis_PRA_90_022307,Chasseur_PRA_31_043421} -- that assume all instances of $g$, $\chi$ and $\del$ to be constant -- lead to errors on the order of $10^{-2}$ (bottom row). Parts (a) and (c) belong to sinusoidal pulses $\Phi_s$, (b) and (d) to tangential ones ($\Phi_t$) given by Eqs.\refeq{fluxpulse}.}
\end{figure}

The parameters we use to simulate the Transmon system yield moderate couplings $g/2\pi\approx 25-30\;{\rm MHz}$ and qubit frequencies on the order of $7\;{\rm GHz}$ around the bias points. We proceed to show that (i) the full TDSWT needs to be applied as soon as one aims at high-fidelity gates and (ii) that the assumption of constant off-diagonal elements severely deteriorates results. The exemplary waveforms we consider are smooth and slow sinusoidal ($\Phi_s$) as well as tangential ($\Phi_t$) controls (as were used in \cite{Chasseur_PRA_31_043421}), both with flux changes of $\del \Phi=60\;{\rm m}\Phi_0$ at maximum, i.e.
\begin{subequations}\label{eq:fluxpulse}\begin{align}
 \Phi_s(t) & = \Phi_{\rm bias} + A\cdot\sin{2\pi\nu t+\varphi}\\
 \Phi_t(t) & = \Phi_{\rm bias} + A\cdot\tan{B\cdot\erf{C\left(t-\frac{t_g}{2}\right)}},
\end{align}\end{subequations}
where $A,B,C,\nu$ and $\varphi$ are constants, $\Phi_{\rm bias}$ is a static bias and $\erf{x}$ is the Gauss error function. We  evolve $\op{\tilde{H}}_{\rm red}$ with these controls and quantify the effects in question by considering the three unitaries
\begin{itemize}
 \item $\op{U}_1$: Full simulation of $\op{\tilde{H}}_{\rm red}$,
 \item $\op{U}_2$: Neglect terms proportional to $\dot{\lambda}$ in $\op{\tilde{H}}_{\rm red}$,
 \item $\op{U}_3$: Neglect terms proportional to $\dot{\lambda}$ and approximate all instances of $g$, $\chi$ and $\del$ as their mean values in $\op{\tilde{H}}_{\rm red}$.
\end{itemize}
To measure the error resulting from those three models, we use the common gate overlap fidelity 
\begin{align}
 F(\op{U}) & = \frac{1}{d_{\mathbb{Q}^2}}\Abs{\Trace{\op{U}^{\dagger}\op{U}_{\rm ideal}}}^2
\end{align}
and choose a set of $N_s=10000$ random unitaries $\op{U}_{\rm ideal}$ of dimension $d_{\mathbb{Q}}=2$, using the representation 
\begin{align}\label{eq:unitary}
		\op{U}_{\rm ideal} = \op{U}(\varphi_1,\varphi_2,\varphi_3) = \begin{pmatrix}e^{i\varphi_1}\cos{\theta} & e^{i\varphi_2}\sin{\theta}\\-e^{-i\varphi_2}\sin{\theta} & e^{-i\varphi_1}\cos{\theta}\end{pmatrix}
\end{align}
for the arbitrary $2\times 2$ unitaries. We evaluate the fidelity for each of the $\op{U}_j$ with respect to the $N_s$ different random target unitaries and compute the fidelity differences
\begin{align}
  \del F(\op{U}_m,\op{U}_n)=F(\op{U}_m)-F(\op{U}_n).
\end{align}
The corresponding normalized histograms are plotted in \reffig{diff1} for sinusoidal (a,c) and tangential (b,d) modulation for realistic gate times of $30\;{\rm ns}$ . 

The top row of \reffig{diff1} reveals that even for smooth pulses without any fast modulation, an incorrect frame transformation (SWT vs. TDSWT) translates into errors in gate fidelities on the order of $10^{-3}$. For models that assume constant off-diagonal components in the reduced Hamiltonian \refeq{dispHamRed}, as was done in earlier studies \cite{Martinis_PRA_90_022307,Chasseur_PRA_31_043421}, the error in gate fidelities is even on the order of $10^{-2}$ (bottom row, \reffig{diff1}). To substantiate the importance of the results, we want to highlight that the pulses we used for simulations are significantly smoother (and free from fast oscillations) than usual optimal control shapes found through e.g. gradient-based optimization routines. Those pulses typically exhibit relatively fast changes, which in turn lead to increasing values for the velocities $\dot{\lambda}_j^{(m)}$ and thereby even higher discrepancies in gate fidelities.

\subsection{Error estimation}
\begin{figure}
 \centering
 \includegraphics[width=.95\linewidth]{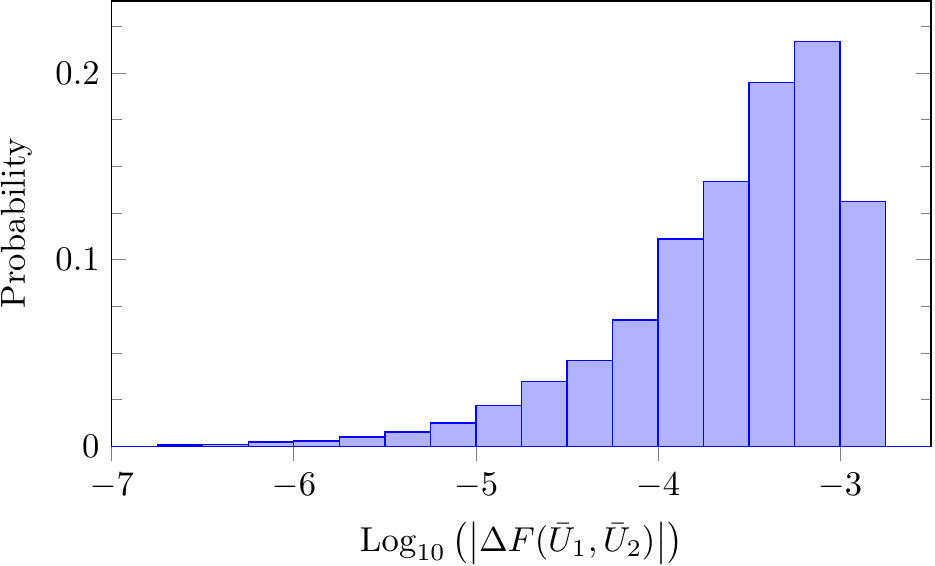}
 \caption{\label{fig:diff2}Normalized histogram for the difference in fidelity $\del F$ with respect to $N_s=10000$ random unitaries obtained from a second order Magnus expansion, see Eq.\refeq{error1}. The statistics are based on a tangential pulse and reproduce those of \reffig{diff1}a very well.}
\end{figure}

A second order Magnus expansion \cite{Warren_JChemPhys_81_5437,Blanes_PhysRep_470_151} can be used to understand the error statistics depicted in \reffig{diff1}. In general, the Magnus expansion is a way to analytically approximate the unitary at time $t_g$ under dynamics of a time-dependent Hamiltonian $\op{H}$ as
\begin{align}
	\op{\bar{U}} & = \exp{-it_g\sum_{k=1}^\infty\op{\bar{H}}^{(k)}}. 
\end{align}
We truncate the series for the unitaries $\op{U}_1$ and $\op{U}_2$ at $k=2$, so that only the first- and second order averaged Hamiltonians
\begin{subequations}\begin{align}
	\op{\bar{H}}^{(1)} & = \hphantom{-}\frac{1}{t_g}\int\limits_0^{t_g}dt\,\op{H}(t)\\
	\op{\bar{H}}^{(2)} & = -\frac{i}{2t_g}\int\limits_0^{t_g}dt_2\int\limits_0^{t_2}dt_1\Comut{\op{H}(t_2)}{\op{H}(t_1)}
\end{align}\end{subequations}
are required. For convenience we introduce the following shorthand notation for Hamiltonian \refeq{dispHamRed}
\begin{subequations}\label{eq:appDef1}\begin{align}
	\omega & = \Braket{11|\op{\tilde{H}}_{\rm red}|11}\\
	g_r & = \hphantom{-}\real{\Braket{11|\op{\tilde{H}}_{\rm red}|20}}\\
	g_i & = -\imag{\Braket{11|\op{\tilde{H}}_{\rm red}|20}}
\end{align}\end{subequations}
which after some standard matrix algebra leads to a closed analytic expression for the error $\del F$, given by
\begin{align}\begin{split}\label{eq:error1}
	\del F(\op{\bar{U}}_1,\op{\bar{U}}_2) = & \hphantom{-}f(k_1,\bar{\omega}+\delta_{g_i,g_r},\delta_{\omega,g_r}-\bar{g}_i,\bar{g}_r+\delta_{\omega,g_i},\vec{\varphi})\\&-f(k_2,\bar{\omega},\delta_{\omega,g_r},\bar{g}_r,\vec{\varphi}).
\end{split}\end{align}
Here, we denote the time-averaged mean of some quantity $s(t)$ with a bar, i.e.
\begin{align}
	\bar{s} & = \frac{1}{t_g}\int\limits_0^{t_g}dt\,s(t).
\end{align}
Information about the unitary's phases enters through the second order Magnus terms, which are determined by the quantities 
\begin{subequations}\label{eq:appDef2}\begin{align}
	\delta_{\omega,g_r} & = \int\limits_0^{t_g}dt_2\int\limits_0^{t_2}dt_1\,\LR{\omega(t_2)g_r(t_1)-\omega(t_1)g_r(t_2)},\\
	\delta_{\omega,g_i} & = \int\limits_0^{t_g}dt_2\int\limits_0^{t_2}dt_1\,\LR{\omega(t_2)g_i(t_1)-\omega(t_1)g_i(t_2)},\\
	\delta_{g_i,g_r} & = \int\limits_0^{t_g}dt_2\int\limits_0^{t_2}dt_1\,\LR{g_i(t_2)g_r(t_1)-g_i(t_1)g_r(t_2)}.
\end{align}\end{subequations}
The rotation angles of unitaries $\op{\bar{U}}_1$ and $\op{\bar{U}}_2$ are set by the constants $k_1$ and $k_2$, respectively. They are given by
\begin{subequations}\label{eq:appDef3}\begin{align}
  k_1 & = \sqrt{(\bar{\omega}+\delta_{g_i,g_r})^2+(\delta_{\omega,g_r}-\bar{g}_i)^2+(\bar{g}_r+\delta_{\omega,g_i})^2},\\
  k_2 & = \sqrt{\bar{\omega}^2+\delta_{\omega,g_r}^2+\bar{g}_r^2},
\end{align}\end{subequations}
Unitary \refeq{unitary} is defined by the angles $\vec{\varphi}=(\varphi_1,\varphi_2,\theta)$. The function $f(k,a_1,a_2,a_3,\vec{\varphi})$ in Eq.\refeq{error1} is defined as
\begin{align}\begin{split}
	f(k,a_1,a_2,a_3,\vec{\varphi})\hspace{16.5em}\\ 
	= \frac{4}{k^2}\!\left\{k\cos{\varphi_1}\!\cos{k}\!\cos{\theta}-\sin{k}\!\left\{a_1\cos{\theta}\!\sin{\varphi_1} \right.\right.\\
	+\left.\left. a_2\cos{\varphi_2}\!\sin{\theta}+a_3\sin{\varphi_2}\!\sin{\theta}\!\right\}\!\right\}^2.\hspace{4em}
\end{split}\end{align}
Indeed, as shown in \reffig{diff2} for tangential pulses, Eq.\refeq{error1} reproduces the statistics of a numerically exact simulation (\reffig{diff1}a) very well. The mean error, independent of $\vec{\varphi}$, is obtained via averaging over $\vec{\varphi}\in\left[0,2\pi\right]^{\otimes 3}$ and yields a value of $\overline{\del F}\sim 10^{-3.21}$ for the case considered in \reffig{diff2}.

%\begin{table}[]
%\centering
%\label{my-label}
%\begin{tabular}{|cc|cc|}
%\hline
%\multicolumn{2}{|c}{Transmon 1}     & \multicolumn{2}{c|}{Transmon 2}     \\ \hline\hline
%$d^{(1)}$                    & 0    & $d_1{(2)}$                    & 0    \\
%$E_c^{(1)}$ (GHz)            & 0.45 & $E_c^{(2)}$ (GHz)            & 0.36 \\
%$C_r^{(1)}/C_{\Sigma}^{(1)}$ & 80   & $C_r^{(2)}/C_{\Sigma}^{(2)}$ & 80   \\
%$\beta^{(1)}$                & 0.28 & $\beta^{(2)}$                & 0.28 \\
%$E_{J\Sigma}^{(1)}$ (GHz)            & 30   & $E_{J\Sigma}^{(2)}$ (GHz)            & 20  \\ \hline
%\end{tabular}
%\caption{\label{tab:params}Transmon parameters used for simulations.}
%\end{table}

%
% -----------------------------------------------------------------------------------
%
\section{Conclusions}
We have given a detailed outline of the time-dependent Schrieffer-Wolff transformation and applied it to derive a general expression for the dispersive Hamiltonian of arbitrarily many multilevel systems coupled to a cavity via a Jaynes-Cummings type of interaction. The usual dispersive coupling between the multilevel systems is altered by terms that scale with the dispersive adiabaticity parameter. 

As a specific example, we provide a simple but accurate model to implement dispersive entangling two-qubit gates using only Z-control of the qubits. Fidelities based on previous models are shown (numerically and analytically) to be off by up to $10^{-2}$ for control fields without fast modulation, which certainly influences high-fidelity gates compatible with scalable fault-tolerant architectures. In the case of high-frequency controls or pulses with fast flux sweeps, one needs to consider higher-order terms of the TDSWT, and gate fidelities based on previous models become even more erroneous. 

As a final note, we want to highlight that the fundamental effects considered in this work are not only apparent in the dispersive frame: For instance the dependence of coupling strengths $g$ on the applied magnetic flux do also impact simulations of the full Jaynes-Cummings Hamiltonian, and should be considered in order to provide accurate simulations of the real dynamics. 
%
% -----------------------------------------------------------------------------------
%
\section{Acknowledgements}
We thank Bruno G. Taketani for useful discussions. The authors acknowledge funding through the LogiQ program of the Intelligence Advanced Research Projects Activity (IARPA) under the grant number W911NF-16-1-0114.
% \ldots and another one\ldots

%----- References ---------
\bibliography{Bibliography.bib}
\bibliographystyle{apsrev4-1}

\end{document}